# Selective Vulnerability to Kainate-induced Oxidative Damage in Different Rat Brain Regions

Eduardo Candelario-Jalil [*], Saied M. Al-Dalain, Rubén Castillo, Gregorio Martínez and Olga Sonia León

*Department of Pharmacology, University of Havana (CIEB-IFAL), Apartado Postal 6079, Havana 10600, CUBA*

[*] Author to whom all correspondence should be addressed:
Eduardo Candelario-Jalil, PhD
Department of Pharmacology
University of Havana (CIEB-IFAL)
Apartado Postal 6079
Havana City 10600
CUBA
Phone: +53-7-219-536
Fax: +53-7-336-811
E-mail: candelariojalil@yahoo.com

**ABSTRACT**

Some markers of oxidative injury were measured in different rat brain areas (hippocampus, cerebral cortex, striatum, hypothalamus, amygdala/piriform cortex and cerebellum) after the systemic administration of an excitotoxic dose of kainic acid (KA, 9 mg/kg, i.p.) at two different sampling times (24 and 48 h). KA was able to markedly ($p<0.05$) lower glutathione (GSH) levels in hippocampus, cerebellum and amygdala/piriform cortex (maximal reduction at 24 h). In a similar way, lipid peroxidation, as assessed by malonaldehyde (MDA) and 4-hydroxyalkenals (4-HDA) levels, significantly increased ($p<0.05$) in hippocampus, cerebellum and amygdala/piriform cortex, mainly at 24 h after KA. In addition, hippocampal superoxide dismutase (SOD) activity significantly decreased ($p<0.05$) with respect to basal levels by 24 h after KA application. On the other hand, brain areas such as hypothalamus, striatum and cerebral cortex seem to be less susceptible to KA excitotoxicity. According to these findings, the pattern of oxidative injury induced by systemically administered KA seems to be highly region-specific. Further, our results have shown that a lower antioxidant status (GSH and SOD) seems not to play an important role in the selective vulnerability of certain brain regions, since it correlate poorly with increases in markers of oxidative damage.

**INTRODUCTION**

Excitotoxicity, the death of neural cells caused by overactivation of postsynaptic glutamate receptors, is considered a key pathological event of neuronal injury in stroke, epilepsy, trauma and neurodegenerative diseases.[1,2]

Kainic acid (KA), a pyrrolidine excitotoxin isolated from the seaweed *Digenea simplex*, is a potent neuroexcitatory compound, which after intracerebral or systemic injection leads to generalised limbic seizures in rats.[3] The neuronal damage induced by systemically administered KA approximates to that seen following repeated temporal lobe seizures[3] and cerebral ischaemia-reperfusion.[4] Therefore, KA provides a valuable tool with which to model some features of ischaemic damage and of the injury induced by repeated epileptic seizures.[5,6]

Current models of KA toxicity support the hypothesis that the main cause of neurotoxicity is the activation of presynaptic KA receptors and the release of endogenous glutamate.[7] The overstimulation of glutamate receptors, with the subsequent marked intracellular calcium ($Ca^{2+}$) rises, has been implicated in the mediation of injury caused by neurotoxins and ischaemia-related insults.[8,9] Intracellular free $Ca^{2+}$ overload may damage the neurones in various ways. Activation of phospholipase $A_2$, phospholipase C, protein kinase C, endonucleases, nitric oxide synthase as well as $Ca^{2+}$-dependent proteases may contribute to the $Ca^{2+}$-mediated cell damage.[9-11] Since many of these enzymatic reactions are free radicals generative,[12] KA-induced reactive oxygen species (ROS) formation may be critical in its excitotoxic effects. In fact, the generation of ROS has been shown to associate with KA neurotoxicity both in vitro[10,13] and in vivo.[14]

According to several histopathological findings, not all brain regions are equally injured following KA administration. Cells are preferentially killed in the amygdala, piriform cortex and the hippocampus.[3,5,6] In a similar way, a previous report has demonstrated a marked reduction in glutathione, a key cellular antioxidant, in hippocampus and amygdala after KA neurotoxic insult.[15] Moreover, several lines of evidence indicate that the neurochemical changes following KA application seem to be highly region-specific.[5,16,17]

Our goal in this study was to determine the pattern of oxidative damage induced by systemically administered KA in selected areas of the rat brain at two different sampling times.

## MATERIALS AND METHODS

### Materials
Kainic acid hydrate, 5,5'-dithio-bis(2-nitrobenzoic acid), pyrogallol and superoxide dismutase (EC 1.15.1.1) were purchased from Sigma Chemical Co. (St. Louis, MO, USA). The Bioxytech LPO-586 kit for lipid peroxidation was obtained from Oxis International (Portland, OR, USA). All other reagents were of the highest quality available.

### Animals
Studies were performed in accordance with the Declaration of Helsinki and with the Guide for the Care and Use of Laboratory Animals as adopted and promulgated by National Institutes of Health (Bethesda, MD, USA). Male Sprague-Dawley rats (CENPALAB, Havana, Cuba) with a body weight of 200-250 g were used for the experiments. The animals were housed in groups of four per cage in a room with controlled 12:12 light/dark cycle and ad libitum food and water.

### Drug treatment and evaluation of behavioural changes
Kainic acid (KA) was given intraperitoneally (n=20) at a dose of 9 mg/kg of body weight (dissolved in saline solution 9 mg/mL), as previously reported.[18,19] Control rats (n=8) were injected with saline (1 mL/kg body weight). Following administration, rats were observed during 4 h for assessment of convulsive behaviour according to the following rating scale described by Sperk et al.[5]: 0: normal, rare wet dog shakes (WDS), no convulsions; 1: intermediate number of WDS, rare focal convulsions affecting head and extremities, staring and intense immobility; 2: frequent WDS, frequent focal convulsions (no rearing or salivation); 3: frequent WDS, progression to more severe convulsions with rearing and salivation (but without falling over); 4: continuous generalised seizures (rearing, falling over), salivation; 5: generalised tonic-clonic seizures, death within 4h in status epilepticus. Only rats showing overt signs of seizures were considered in this study (n=17), as previously reported.[18,19]

### Sample preparation for biochemical analyses
After 24 h (n=9) and 48 h (n=8) of KA administration, animals were anaesthetised with diethyl ether and perfused intracardially with ice-cold saline in order to eliminate the excess of iron (bound to haemoglobin) that may artificially increase free radical formation. Brains were quickly removed, kept in ice-cold saline and immediately dissected on a cold plate using the atlas of Paxinos and Watson[20] as a reference. Six different regions (hippocampus, cerebral cortex, striatum, cerebellum, hypothalamus and amygdala/piriform cortex) were removed, weighed and homogenised in 20 mM Tris(hydroxymethyl)-aminomethan-HCl buffer (pH 7.4) according to the homogenisation procedure described by Hall et al.[21] Briefly, each brain region was placed in a 2 mL microcentrifuge tube containing rupture glass balls (1 mm diameter) and 1 mL of buffer. Tissues were minced and vortexed for 2 min and then centrifuged at 12 000 rpm during 5 min. The supernatant was collected and employed for biochemical analyses.

**Lipid peroxidation assay**
Lipid peroxidation was assessed by measuring the concentration of malonaldehyde (MDA) and 4-hydroxyalkenals (4-HDA) using the Bioxytech LPO-586 kit. The assay was conducted according to the manufacturer's instructions. This kit takes advantage of a chromogenic reagent (N-methyl-2-phenylindole) which reacts with MDA and 4-HDA at 45°C yielding a stable chromophore with maximal absorption at a wavelength of 586 nm. This wavelength and the low temperature of incubation used in this procedure minimise interferences and undesirable artifacts.[22]

**Glutathione (GSH) determination**
Reduced glutathione was determined spectrophotometrically in the deproteinised samples according to the procedure of Ellman and Lysko.[23]

**Superoxide Dismutase (SOD) activity**
Total SOD activity was measured using pyrogallol as substrate.[24] This method follows the superoxide driven auto-oxidation of pyrogallol at pH 8.2 in the presence of ethylenediaminetetraacetic acid (EDTA). The standard assay mixture contained 1 mM EDTA in 50 mM Tris(hydroxymethyl)-aminomethan-HCl buffer (pH 8.2) with or without the sample. The reaction was started by the addition of pyrogallol (final concentration 0.124 mM) and the oxidation of pyrogallol was followed for 1 min at 420 nm. The percent inhibition of the auto-oxidation of pyrogallol by SOD present in the tissue sample was determined, and standard curves using known amounts of purified SOD (Sigma) under identical conditions were established. One unit of SOD activity was defined as the amount that reduced the absorbance change by 50%, and results were expressed as Units/mg protein.

**Protein determination**
Total protein concentration was determined by the method of Bradford[25] with bovine serum albumin as standard.

**Statistical analysis**
Data are expressed as mean values ± SD (standard deviations). Results were analysed by one-way analysis of variance (ANOVA). If the *F* values were significant, the Student-Newman-Keuls post-hoc test was used to compare groups. In order to determine the relation between antioxidant status and increases in lipid peroxidation, a correlation analysis was performed by determining Pearson's correlation coefficient. Statistical significance was accepted for $P<0.05$ and $P<0.01$.

RESULTS

**Behaviour**
No behavioural changes were observed in vehicle-treated animals. There were considerable differences in the behavioural response in the rats treated with KA (9 mg/kg, i.p.). Seventeen of the 20 animals injected with KA developed tonic-clonic forelimb movements, a main clinical feature of KA neurotoxicity. Within 40-60 min after KA application, the convulsive activity progressed to generalised limbic seizures. Although the intensity of these symptoms showed considerable inter-individual variation, the average symptom rating in KA group reached a value of 3 in Sperk's scale.

**Oxidative damage induced by in vivo KA excitotoxicity**
As shown in Fig. 1, the systemic administration of an excitotoxic dose of KA (9 mg/kg, i.p.) produced a marked decrease in GSH content in hippocampus, amygdala/piriform cortex and cerebellum with respect to basal levels, mainly at 24 h. Although GSH content was significantly reduced in the hippocampus and cerebellum by 48 h as compared with control group ($p<0.05$), it was greater than that by 24 h ($p<0.05$). This finding may suggest that a recovery process in GSH levels is taking place in the regions affected by KA at this time point. In contrast, no statistically significant differences in the content of GSH were observed in the cerebral cortex, striatum and hypothalamus at any time point, as depicted in Fig. 1.

In a similar way, lipid peroxidation, as assessed by MDA and 4-HDA levels, showed a significant increase in hippocampus, amygdala/piriform cortex and cerebellum ($p<0.05$), mainly at 24 h, as shown in Fig. 2. Only the regions in which GSH content was statistically reduced, showed a marked increase in lipid peroxidation endproducts. As judged by the depletion in GSH content and the increase in lipid peroxidation, oxidative injury at 48 h is lower than that by 24 h (Fig. 1 and 2).

Further, we measured the activity of SOD, a key antioxidant enzyme, in each brain region under study. As seen in Table 1, basal SOD activity showed significant differences among the brain areas, being lower in hypothalamus and striatum. At 24 h following KA administration, hippocampal SOD activity was significantly reduced in comparison with the control group ($p<0.05$). No differences were found in other brain regions at any time point after KA (Table 1).

Table 2 shows the results of the correlation analysis between increases in lipid peroxidation and antioxidant status (GSH and SOD) in each brain area. No significant correlation's coefficients were found in any of the regions examined at 24 h (maximal oxidative injury).

DISCUSSION

Our finding that the intraperitoneal administration of KA produces inconsistent behavioural responses is consistent with previous reports.[26,27,28] The exact mechanisms underlying the considerable differences in the response to the behavioural changes by KA in rats are currently unknown. However, it is known that KA poorly penetrates the blood-brain barrier (< 1% of the systemically injected compound can reach the brain)[29] and there can be differences between individual animals in the anatomy and/or physiology of the blood-brain barrier. Therefore, the poor bioavailability of KA, coupled with differences in the properties of the blood-brain barrier, may contribute to the significant differences in the behavioural response produced by the systemic administration of KA.

Neurotoxicity in acute as well as chronic neurological diseases may be partly mediated by oxidative stress caused by overactivation of glutamate receptors.[1] Brain cells are particularly prone to free radical damage because of their high content of iron and polyunsaturated fatty acids, the latter being a substrate for lipid peroxidation, and because of their relatively deficient antioxidative defence mechanisms.[30] The ability of KA to induce oxidative damage has been well documented in the literature.[10,31,32] It is generally accepted that the overactivation of excitatory amino acid receptors triggers marked intracellular $Ca^{2+}$ rises and consequent oxygen radical production. The generation of free radicals by KA and its correlation with excitotoxicity have been proposed by several groups.[4,6,10]

Our present results revealed that the systemic administration of KA was able to induce oxidative damage in particular areas of the rat brain mainly at 24 h. The most vulnerable areas to in vivo KA-mediated oxidative stress resulted to be the hippocampus, cerebellum and amygdala/piriform cortex, which is very similar to the pattern of neuronal loss assessed histopathologically.[3,5] Taking into account our findings, it seems that hypothalamus, striatum and cerebral cortex are resistant to KA-induced oxidative injury.

The mechanisms underlying this highly region-specific pattern of oxidative damage are far from being well understood. In an attempt to explain, at least partially, this selective pattern of oxidative damage associated to KA excitotoxicity, we speculated that certain neuronal populations may be more vulnerable to oxidative stress as a result of a greater oxidative burden or, alternatively, lower antioxidant protection. We have found that the levels of two important neuronal antioxidants (GSH and SOD) correlate poorly with increases in lipid peroxidation in all of the brain areas that were examined (Table 2). For example, the hippocampus was the most oxidatively injured area in spite of showing a relatively high antioxidant capacity with respect to other areas before KA application (Table 1 and Fig. 1).
Therefore, other mechanisms could account for the selective vulnerability to excitotoxic insult, such as the heterogeneous distribution of KA receptors in the rat central nervous system. According to a recent report, KA receptors are expressed in a high density in the pyramidal neurones of the CA1-CA3 subfields of the hippocampus, amygdala, entorhinal and piriform cortices and cerebellar Purkinje cells.[33] Coincidentally, these areas have shown a greater susceptibility to KA-mediated oxidative injury, as observed in Fig. 1 and 2.

Basal SOD activity showed significant differences among the regions under study (Table 1). It seems that neuronal populations with a high density of excitatory amino acids receptors[33,34] display higher SOD activity in view of their predisposition to excitotoxic damage and subsequent free radical production. According to our results, hippocampal SOD activity significantly decreased with respect to vehicle-treated group by 24 h after KA administration (Table 1). It was recently found that the intrastriatal injection of glutamate receptor subtype agonists, such as N-methyl-D-aspartate or KA, significantly reduced Cu,Zn-SOD activity with respect to basal levels.[35] On the other hand, it has been shown that under oxidative stress conditions, the overproduction of free radicals is able to inhibit SOD activity.[36] This latter evidence might explain our finding of reduction of SOD mediated by KA. In addition, the generation of large amounts of nitric oxide, an event associated to excitotoxic damage,[10, 14] decreases the activity and the protein levels of SOD.[37]

The mechanism whereby KA causes intracellular GSH content to fall is not fully understood. A major endogenous protective system is the GSH redox cycle: GSH acts both as a nucleophilic scavenger of numerous compounds and as a cofactor for glutathione peroxidase and glutathione-S-transferase.[30] These two important roles for GSH might explain its depletion under KA-induced oxidative stress conditions. Further, oxygen radicals have been reported to inactivate glutathione reductase.[38]

In the present study, SOD activity and GSH levels increased within 48 h in the susceptible areas following KA excitotoxic insult (Fig. 1 and Table 1). These findings suggest that cells respond to oxidative stress by increasing antioxidative mechanisms. However, the observation that SOD activity and GSH levels increase several hours after the generation of free radicals by KA indicates that these antioxidants are either slow to respond to changes in oxygen radicals levels, or that the levels of free radicals must increase to some threshold level in order to signal an increase in antioxidant defences.

In summary, the results of this study indicate that the events associated with KA-mediated oxidative damage occur in specific brain areas that seem to be highly susceptible to excitotoxic insult. Our findings have shown that a lower antioxidant level seems not to play a role in the selective vulnerability of certain brain regions, since it does not correlate with increases in markers of oxidative damage. Further studies are needed to determine the status of other antioxidant mechanisms in different regions of the rat brain, as well as other metabolic factors that may predispose selected neuronal populations to oxidative damage.

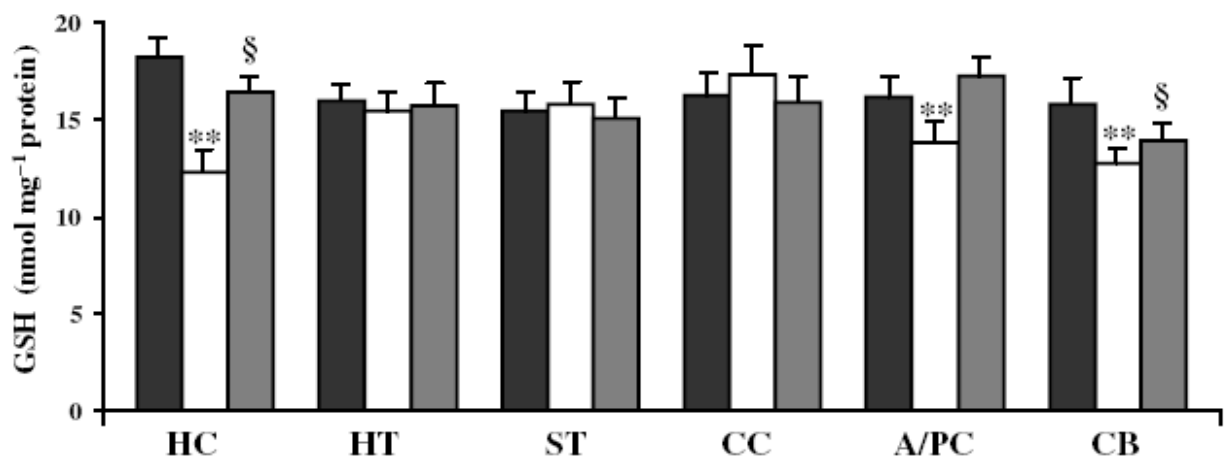

Figure 1. Glutathione (GSH) levels in each brain area after 24 h (open bars) and 48 h (grey bars) of kainic acid (KA) administration (9 mg kg$^{-1}$ i.p.); **$P < 0.01$ with respect to basal levels; (black bars) §$P < 0.05$ with respect to control and KA after 24 h (ANOVA followed by Student–Newman–Keuls *post hoc* test). Values represent means ±SD. HC = hippocampus; HT = hypothalamus; ST = striatum; CC = cerebral cortex; A/PC = amygdala/piriform cortex; CB = cerebellum.

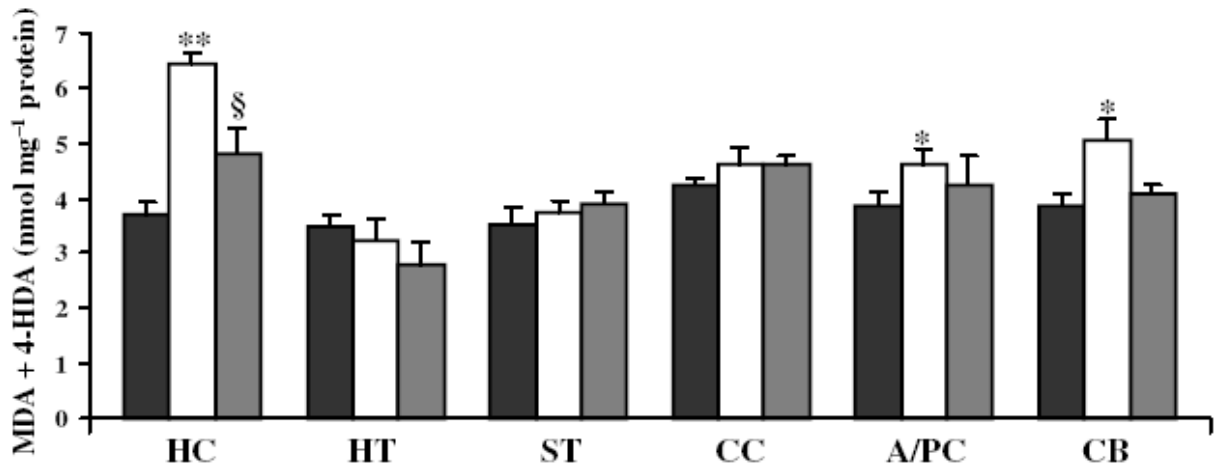

**Figure 2.** Lipid peroxidation as assessed by malonaldehyde (MDA) and 4-hydroxyalkenal (4-HDA) levels after 24 h (open bars) and 48 h (grey bars) of KA application (9 mg kg$^{-1}$ i.p.); *$P < 0.05$ and **$P < 0.01$ with respect to basal levels; §$P < 0.05$ with respect to control (black bars) and KA after 24 h (ANOVA followed by Student–Newman–Keuls *post hoc* test). Values represent means ±SD. HC = hippocampus; HT = hypothalamus; ST = striatum; CC = cerebral cortex; A/PC = amygdala/piriform cortex; CB = cerebellum.

Table 1—Superoxide dismutase (SOD) activity in each brain region and effects of kainic acid (KA) administration

| Brain area | SOD activity (U mg$^{-1}$ protein) | | |
|---|---|---|---|
| | Basal levels | KA 24 h | KA 48 h |
| Cerebellum | $5.3 \pm 1.6^c$ | $5.4 \pm 1.5$ | $6.2 \pm 2.0$ |
| Hypothalamus | $3.8 \pm 0.7^b$ | $3.3 \pm 0.9$ | $4.1 \pm 0.5$ |
| Hippocampus | $8.7 \pm 0.9^a$ | $2.9 \pm 1.2^*$ | $7.8 \pm 1.5$ |
| Amygdala/piriform cortex | $5.1 \pm 1.6^c$ | $4.4 \pm 0.7$ | $5.2 \pm 0.5$ |
| Striatum | $3.2 \pm 0.2^b$ | $3.8 \pm 0.5$ | $3.5 \pm 0.6$ |
| Cerebral cortex | $6.1 \pm 0.8^c$ | $6.0 \pm 1.3$ | $5.6 \pm 0.9$ |

* Values represent means $\pm$SD; $P < 0.05$ with respect to basal levels. Values with non-identical superscripts are significantly different ($P < 0.05$) within the same set.

Table 2—Correlation analysis using Pearson's correlation coefficient $R$ between antioxidant status (GSH and SOD) and increases in lipid peroxidation (MDA + 4-HDA) in each brain region at 24 h (maximal oxidative injury) ($n = 9$)

| Brain region | MDA+4-HDA vs. GSH levels | | MDA+4-HDA vs. SOD | |
|---|---|---|---|---|
| | R | P | R | P |
| Hippocampus | −0.05 | 0.97 | 0.53 | 0.64 |
| Hypothalamus | 0.57 | 0.61 | 0.68 | 0.52 |
| Striatum | 0.38 | 0.75 | −0.14 | 0.91 |
| Cerebral cortex | −0.79 | 0.41 | −0.87 | 0.33 |
| Amygdala/piriform cortex | 0.26 | 0.83 | −0.83 | 0.38 |
| Cerebellum | −0.79 | 0.42 | 0.04 | 0.98 |